\input epsf
\documentstyle[sprocl]{article}
\newcommand{\bdis}{\begin{displaymath}}
\newcommand{\edis}{\end{displaymath}}
\newcommand{\beqn}{\begin{equation}}
\newcommand{\eeqn}{\end{equation}}
\newcommand{\beqna}{\begin{eqnarray}}
\newcommand{\eeqna}{\end{eqnarray}}

\newcommand{\eps}{\varepsilon}
\newcommand{\del}{\partial}
\newcommand{\dcsb}{D$\chi$SB}
\def\AJP(#1,#2,#3){Aust. J. Phys. {\bf #1} (19#2) #3}
\def\NP(#1,#2,#3){Nucl. Phys. {\bf #1} (19#2) #3}
\def\NPBPS(#1,#2,#3){Nucl. Phys. {\bf B}(Proc. Suppl.){\bf #1} (19#2) #3}
\def\PL(#1,#2,#3){Phys. Lett. {\bf #1} (19#2) #3}
\def\PR(#1,#2,#3){Phys. Rev. {\bf #1} (19#2) #3}
\def\PRL(#1,#2,#3){Phys. Rev. Lett. {\bf #1} (19#2) #3}
\def\PTP(#1,#2,#3){Prog. Theor. Phys. {\bf #1} (19#2) #3}
\def\PTPS(#1,#2,#3){Prog. Theor. Phys. (Suppl.){\bf #1} (19#2) #3}

\newcommand{\conf}{{\rm conf}}

\newcommand{\Yu}{{\rm Y}}
\newcommand{\Cou}{{\rm C}}

\newcommand{\gpar}{\alpha_{\rm e}}
\newcommand{\Nf}{N_{\rm f}}
\newcommand{\Nc}{N_{\rm c}}

\begin{document}

%
\title
{Properties of the QCD vacuum \\
in the dual Ginzburg-Landau theory}
\author
{S.~Umisedo, H.~Toki and H.~Suganuma}
\address{Research Center for Nuclear Physics, Osaka University}
\maketitle
\abstract
{We study the properties of the QCD vacuum in the dual Ginzburg-Landau theory,
where QCD-monopole condensation leads to the linear confinement potential.
Using the effective potential, we find monopole dominance for chiral-symmetry
breaking.
The effective mass of off-diagonal (charged) gluons is estimated as
$M_{ch} \simeq 1.31$ GeV.
The gluon condensate is calculated as
$\langle\frac{\alpha_{\rm s}}{\pi}G^2 \rangle$
$=$
$(344 {\rm MeV})^4$.
}
%
%
%
\vspace{3mm}
Recent studies of the lattice gauge theory
strongly support the dual superconductor
picture for the QCD vacuum
in the confinement phase in the 't~Hooft abelian gauge
	\cite{thooft},
and show abelian dominance and monopole dominance 
for the confinement and dynamical chiral-symmetry breaking (D$\chi$SB)
	\cite{miyamura}.
The dual Ginzburg-Landau (DGL) theory is an infrared effective theory of QCD
based on dual Higgs mechanism in the abelian gauge
	\cite{suganumaA,sasaki},
and is described by the diagonal gluon
$A^\mu$,
the dual gauge field
$B^\mu$,
and the QCD-monopole field $\chi_\alpha$
($\alpha = 1,2,3$) 
	\cite{suganumaA}.

In the QCD-monopole condensed vacuum
$|\chi_\alpha|^2 \equiv v^2$,
the dual gauge field acquires a mass of
$m_B = 4\pi\sqrt{3}v/e$
and propagates as a massive vector field
	\cite{suganumaA}.
On the other hand, the nonperturbative gluon propagator takes the form 
$D_{\mu\nu}(p)$
$=$
$-d(p^2)$
$T_{\mu\nu}(p)$
$/p^2$
$-$
$\gpar$
$L(p)_{\mu\nu}$
$/p^2$,
where $T_{\mu\nu}=g_{\mu\nu}-L_{\mu\nu}$ and $L_{\mu\nu}=p_\mu p_\nu /p^2$ 
is the transverse and the longitudinal tensor, respectively.
In the DGL theory, $d(p^2)$ is given as
\begin{equation}
d(p^2)  = \frac{1}{3}
	  \left\{
	  {4m_B^2 \over (p^2+m_B^2)} \cdot
	  {p^2+a^2 \over a(a + \sqrt{a^2 + p^2})}
	+ {2p^2 \over p^2+m_B^2} + 1
	  \right\}
\label{eqn:GpropTr}
\end{equation}
with the infrared cutoff parameter $a$ corresponding to the hadron size 
introduced in relation with the polarization effect of light quarks
	\cite{sasaki}.
The first and the second term, $d_\conf(p)$ and $d_\Yu(p)$ is related to the
linear confinement and Yukawa part in the static quark potential, respectively.
The third term, $d_\Cou(p)$ does not contribute to the static potential.

First, we discuss \dcsb~ in the quark sector physics.
Our group showed the essential role of QCD-monopole condensation 
to D$\chi $SB by solving the Schwinger-Dyson (SD) equation 
	\cite{suganumaA,sasaki}. 
For further investigation, we calculate the effective potential
to extract the contribution of the confinement effect energetically.
Using the nonperturbative gluon propagator
in the DGL theory, the effective potential is expressed as
	\cite{umisedo}
\begin{eqnarray}
\hspace{-5mm}{}& &\hspace{-3mm}V_{\rm eff}[M(p^2)]
	=
	-   2\Nc \Nf \int\!\! {d^4p \over (2\pi)^4}
	  \{
	   {\rm ln}({p^2+M^2(p^2) \over p^2})
	   -  2{M^2(p^2) \over p^2+M^2(p^2)}
	   \}
\nonumber \\
\hspace{-5mm}	&+&\hspace{-3mm}
	    \Nf (\Nc -1) \int\!\! 
	    {d^4p \over (2\pi)^4} {d^4q \over (2\pi)^4}
	    e^2 {M(p^2) \over p^2+M^2(p^2)} 
	   {M(q^2) \over q^2+M^2(q^2)} 
	    D_{\mu \mu }(p-q)  \label{eqn:EffPot} 
\end{eqnarray}
in the improved ladder approximation
	\cite{higashijima}.
The first term is the quark-loop contribution. 
The second term 
is divided into three parts 
($V_{\rm conf}$,$V_\Yu$,$V_\Cou$) 
corresponding to the decomposition of $D_{\mu \mu }$ in 
	Eq.(\ref{eqn:GpropTr}).

Fig.1 shows $V_{\rm eff}[M(p^2)]$ as a function 
of the infrared quark mass $M(0)$ using the quark-mass ansatz 
which is consistent with 
the renormalization group analysis of QCD
	\cite{higashijima}.
The parameters $e = 5.5$, $m_B = 0.5$ GeV and $a = 85$ MeV
are chosen so as to reproduce the inter-quark potential and 
the flux-tube radius
$ R \simeq $0.4 fm
	\cite{suganumaA}.
One finds a nontrivial minimum at $M(0) \simeq $0.4 GeV in 
$V_{\rm eff}[M(p^2)]$.
Fig.2 shows the separate contribution 
$V_{\rm quark}$, $V_{\rm conf}$, $V_\Yu$ and $V_\Cou$.
The effective potential is mainly lowered by $V_{\rm conf}$.
Since the lowering of the effective potential contributes to D$\chi $SB, 
the main driving force of D$\chi $SB is brought by the confinement part 
$D_{\mu \mu }^{\rm conf}(p)$ 
in the nonperturbative gluon propagator, 
which means monopole dominance for D$\chi $SB 
	\cite{miyamura}.\\
%
%
\noindent
%
\begin{minipage}[ht]{47.5mm}
\epsfxsize=47.5mm
\epsfbox{figure1.psn}

\noindent
\baselineskip 9pt
{\footnotesize
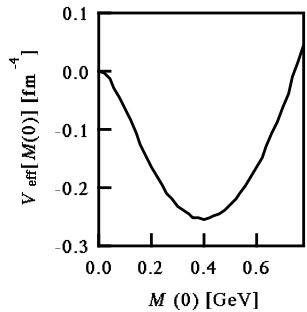
Fig.1 : The effective potential $V_{\rm eff}[M(p^2)]$ as a
function of infrared quark mass $M(0)$.
}
\end{minipage}
%
\hspace{5mm}
%
\begin{minipage}[ht]{47.5mm}
\epsfxsize=47.5mm
\epsfbox{figure2.psn}

\noindent
\baselineskip 9pt
{\footnotesize
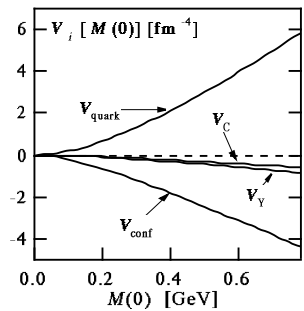
Fig.2 : The separation of the effective potential.
The confinement part $V_\conf$ 
dominates $V_\Yu$ and $V_\Cou$.
}
\end{minipage}
%
%

\vspace{10pt}
Next, we discuss the gluon sector physics.
The observed feature of abelian dominance for confinement and \dcsb~
indicates that the remaining off-diagonal gluon
$C^\alpha_\mu$
($\sqrt{2}C^1_\mu = A^1_\mu + iA^2_\mu$, {\em etc.}),
which behaves as charged matter field in the abelian gauge,
is irrelevant to the low-energy physics.
Here, we study the self-energy of the charged gluon
using the SD equation for $C^\alpha_\mu$ in the DGL theory.
In the Hartree-type approximation,
the SD equation for $C^\alpha_\mu$ is expressed as
\newpage
\beqn
i\Delta^M_{\mu\nu}
	=   i\Delta^0_{\mu\nu}
	  + \frac{1}{2} i\Delta^M_{\mu\rho} 
	  		 \Gamma^{ab}_{\rho\sigma\tau\upsilon}
	  		iD^{ab}_{\sigma\tau}
	  		i\Delta^0_{\upsilon\nu},
\label{eqn:SD}	  		
\eeqn
where the full propagator of $C^\alpha_\mu$ is assumed to be
$i\Delta^M_{\mu\nu}$
$=$
$-iT_{\mu\nu}$
$/(k^2-M^2)$
in the Landau gauge, while the free one $i\Delta^0_{\mu\nu}$ and the
vertex factor
$\Gamma^{ab}_{\kappa\lambda\mu\nu}$
$	= - ie^2\eps^a_\alpha \eps^b_\alpha$
$	    \{ 2g_{\kappa\nu}g_{\lambda\mu}$
$	    	   - g_{\kappa\lambda}g_{\mu\nu}$
$	    	   - g_{\kappa\mu}g_{\lambda\nu}\}$
is obtained from the QCD Lagrangian.
The self-energy $\Sigma$ defined by 
$\Sigma_{\rho\upsilon}$
$\equiv$
$\Gamma^{ab}_{\rho\sigma\tau\upsilon}$
$D^{ab}_{\sigma\tau}/2$
$=$
$g_{\rho\upsilon}\Sigma$
is related to the effective mass $M$ of the charged gluon as $\Sigma = M^2$.
To drop off the divergence from the perturbative part in
$d(p^2)$,
we replace
$d(p^2)$ by
$\tilde{d}(p^2)$
$\equiv$
$d(p^2;m_B)$
$-$
$d(p^2;0)$
in the momentum integration in Eq.(\ref{eqn:SD}).
Then, the self-energy is given by the integral
$\Sigma$
	$=$  $\frac{9e^2}{4}$
	   $\int \frac{d^4p}{(2\pi)^4}$
	   $\frac{\tilde{d}(p^2)}{p^2}$.
Considering the reduction of the monopole condensate in the core region
of the dual Abrikosov vortex
	\cite{suganumaA,umisedo},
we introduce a physical ultraviolet cutoff to the
monopole condensate as
$m_B(p)$
$=$
$m_B \theta(m_\chi - p)$.
We estimate the charged gluon mass as
$M_{ch}$
$=$
1.31 GeV, which suggests the irrelevance of the charged gluon
for low-energy physics in the abelian gauge.
This property is also seen in the lattice QCD
	\cite{amemiya}.

Finally, we investigate the gluon condensate.
With the assumption of abelian dominance for the gluon condensate
$\langle G^a_{\mu\nu} G_a^{\mu\nu} \rangle$
	$\simeq$
		$\langle \vec{f}_{\mu\nu} \vec{f}^{\mu\nu} \rangle$,
where $\vec{f}_{\mu\nu}$ is the abelian field strength tensor
	\cite{suganumaA},
we can estimate it within the DGL theory.
Since the non-perturbative QCD vacuum is filled with condensed monopoles,
it is convenient to proceed in the dual field formalism,
$\langle \vec{f}_{\mu\nu} \vec{f}^{\mu\nu} \rangle$
	$=$  $-$  $\langle {}^\ast\vec{f}_{\mu\nu}
			{}^\ast\vec{f}^{\mu\nu}	\rangle$
	$=$  $-$  $2( g^{\alpha\beta} g^{\gamma\delta}$
		  $-g^{\beta\gamma} g^{\delta\alpha})$
		$\del^x_\alpha \del^y_\beta$
		$\langle \vec{B}_\gamma(x)
			\vec{B}_\delta(y) \rangle |_{x=y}$,
to obtain the expression
\beqn
\langle \vec{f}^2 \rangle
	=	12i\int\frac{d^4k}{(2\pi)^4}
			\frac{k^2}{k^2 - m_B^2(k)}.
\eeqn
By subtracting the contribution of perturbative
part,
$\langle \vec{f}^2 \rangle_{m_B}$
$-$
$\langle \vec{f}^2 \rangle_{m_B = 0}$,
we obtain the value
$\langle\frac{\alpha_{\rm s}}{\pi}G^2 \rangle$
$=$
$(344{\rm MeV})^4$,
which is comparable to the values obtained by other studies
	\cite{narison,campostrini}.

%
%
%
%


\small
\baselineskip 10pt
\begin{thebibliography}{99}
\parskip 0pt
%
%
%
%
%
\bibitem{thooft}\label{bib.4}
G.~'t~Hooft, Nucl.~Phys.~{\bf B190} (1981) 455.
%
\bibitem{miyamura}\label{bib.5}
O.~Miyamura, Nucl.~Phys.~{\bf B}(Proc.~Suppl.){\bf 42} (1995) 538.
%
%
%
\bibitem{suganumaA}\label{bib.8}
H.~Suganuma, S.~Sasaki and H.~Toki, 
Nucl.~Phys.~{\bf B435} (1995) 207.
%
\bibitem{sasaki}\label{bib.10}
S.~Sasaki, H.~Suganuma and H.~Toki, 
Prog.~Theor.~Phys.~{\bf 94} (1995) 373;
\PL(B387,96,145).
%
%
\bibitem{umisedo}\label{bib.9}
S.~Umisedo, H.~Suganuma and H.~Toki,
Phys. Rev. {\bf D} Feb. 1.\\
H.~Suganuma, S.~Umisedo, S.~Sasaki, H~.Toki and O.~Miyamura,
\AJP(50,97,233).
%
%
\bibitem{higashijima}\label{bib.12}
K.~Higashijima, Prog.~Theor.~Phys.(Suppl.) {\bf 104} (1991) 1.
%
\bibitem{amemiya}
K.~Amemiya and H.~Suganuma,
this Conference.
%
\bibitem{narison}
S.~Narison,
\PL(B387,96,162).
%
\bibitem{campostrini}
M.~Campostrini et al.,
\PL(B225,89,393).
%
%
\end{thebibliography}
\end{document}